\documentclass[english,11pt,a4paper]{article}
\usepackage[T1]{fontenc}
\usepackage[latin9]{inputenc}
\usepackage{amsmath}
\usepackage{amsthm}
\usepackage{amssymb}
\usepackage{graphicx}

\usepackage{ulem}

\makeatletter

\numberwithin{equation}{section}
\numberwithin{figure}{section}

\usepackage{jheppub}

\usepackage{tikz}
\usetikzlibrary{shapes,arrows}

\makeatother

\usepackage{babel}
\begin{document}
	\preprint{ULB-TH/21-13}
	\title{Probing the muon $g-2$ with future beam dump experiments}
	
	\abstract{
		We consider the light $Z'$ explanation of the muon $g-2$ anomaly. 
		Even if such a $Z'$ has no tree-level coupling to electrons, in general one will be induced at loop-level. 
		We show that future beam dump experiments are powerful enough to place stringent constraints on\textemdash or discover\textemdash a $Z'$ with loop-suppressed couplings to electrons. 
		Such bounds are avoided only if the $Z'$ has a large interaction with neutrinos, in which case the scenario will be bounded by ongoing neutrino scattering experiments. 
		The complementarity between beam dump and neutrino scattering experiments therefore indicates that there are good prospects of probing a large part of the $Z'$ parameter space in the near future. 
	}
	
	\author{Rupert Coy,}
	\author{Xun-Jie Xu}
	\affiliation{Service de Physique Th\'{e}orique, Universit\'{e} Libre de Bruxelles, Boulevard du Triomphe, CP225, 1050 Brussels, Belgium}
	\maketitle
	
	\section{Introduction}
	Following the new measurement of the anomalous magnetic moment of the muon at Fermilab, which has pushed the tension between theory and experiment to $4.2\sigma$ \cite{Muong-2:2006rrc,Muong-2:2021ojo,Aoyama:2020ynm}, the $(g-2)_\mu$ anomaly must be considered one of the most compelling indications of new physics.\footnote{Notwithstanding a lattice calculation which suggests a much smaller discrepancy \cite{Borsanyi:2020mff}.} 
	Copious explanations have been proposed in the literature, covering a broad range of possibilities. 
	One notable class of solutions involves extending the Standard Model (SM) gauge sector by an additional $U(1)'$ symmetry, whose gauge boson, $Z'$, is responsible for the shift in $(g-2)_\mu$ \cite{Pospelov:2008zw}. 
	Not only is this scenario very simple,
	involving only one extra field, it could potentially be detectable at a large number of different experiments, depending on the mass and couplings of the $Z'$.

	Basic $Z'$ models such as the dark photon, in which the gauge boson couples to SM fermions only via kinetic mixing \cite{Holdom:1985ag}, have been ruled out as a solution to the $(g-2)_\mu$ anomaly. 
	This is due to a combination of collider, beam dump, astrophysical and cosmological bounds, see e.g. \cite{Ilten:2018crw,Bauer:2018onh}. 
	The majority of these bounds are on $Z'$ interactions with electrons, neutrinos and light quarks, while its couplings to second and third generation fermions are relatively less constrained. 
	Consequently, attention has shifted to frameworks of a $Z'$ with flavour-dependent interactions, for instance a gauged $L_{\mu}-L_{\tau}$ symmetry wherein the $Z'$ does not couple to electrons or quarks at tree-level \cite{He:1990pn,Foot:1990mn,He:1991qd}. 
	These types of models permit the necessary $Z'$ coupling to muons to explain the $(g-2)_\mu$ anomaly while at the same time seemingly avoid many of the most severe experimental constraints, see e.g. \cite{Altmannshofer:2014pba,Altmannshofer:2016brv,Escudero:2019gzq,Garani:2019fpa,Bodas:2021fsy}. 
	The preferred parameter space for these secenarios is $m_{Z'} < 2m_\mu$, in which case the $Z'$ cannot decay to muons and so avoids powerful BaBar limits \cite{BaBar:2016sci}.

	In reality, the situation is not so straightforward. 
	Even if the $Z'$ does not couple to electrons or quarks at tree level, an effective coupling will be generated at loop level. 
	Despite the loop suppression, there remain a variety of important constraints, which may also depend sensitively on the size of the $Z'$ couplings to neutrinos. 
	In the limit of feeble $Z'$-$\nu$ interactions, 
	there are powerful bounds on $Z'$-electron couplings down to $\mathcal{O}(10^{-8})$~\cite{Bauer:2018onh}. 
	Clearly, this is relevant even for loop-induced couplings. 
	This is due in particular to results from historical beam dumps experiments (see e.g. \cite{Andreas:2012mt}) and forecasts by ongoing and future ones such as NA64 \cite{NA64:2019auh}, SHiP \cite{Alekhin:2015byh,SHiP:2015vad}, SeaQuest \cite{Gardner:2015wea,Berlin:2018pwi,Tsai:2019buq}, MATHUSULA \cite{Chou:2016lxi,Evans:2017lvd}, FASER \cite{Feng:2017uoz,FASER:2018eoc,Abreu:2019yak}, and CODEX-b \cite{Gligorov:2017nwh}. 
	These bounds become relatively weaker if the $Z'$ has sizeable interactions with neutrinos. 
	In that case, however, neutrino scattering experiments provide competitive constraints on the $Z'$ \cite{Bilmis:2015lja,Lindner:2018kjo,Abdullah:2018ykz,Link:2019pbm,Ballett:2019xoj,Dev:2021xzd}. There is thus a complementarity between BD and neutrino experiments for probing the $Z'$ solution to the muon $g-2$ anomaly.
	
	The potentially important role of neutrino couplings in BD experiments and the aforementioned complementarity were largely overlooked in the past, and constitute the main focus of the paper.

	This work is organized as follows: in Sec.~\ref{sec:Framework}, we establish the conventions for the analysis, including the effects of kinetic mixing, mass mixing and loop diagrams on $Z'$ couplings to fermions. 
	We then outline the present experimental status of the $Z'$ explanation of the $(g-2)_\mu$ anomaly. 
	The focus turns to future beam dump experiments in Sec.~\ref{sec:BD}. 
	After reviewing the physics of beam dumps, we demonstrate how the bounds depend sensitively on the $Z'$ couplings to both electrons and neutrinos. 
	In Sec.~\ref{sec:results}, we combine present and expected future bounds, finding that most of the currently viable parameter space will be probed by a combination of future BD and neutrino experiments.
	Fig.~\ref{fig:four-plot} summarises our key findings, and we conclude in Sec.~\ref{sec:con}.
	
	\section{Framework}
	\label{sec:Framework}
	
	\subsection{Generic $Z'$ couplings}
	\label{subsec:generic}
	
	We consider a  generic framework in which a weakly-coupled, neutral vector boson
	$Z'$, typically originating from hidden $U(1)'$ extensions of the
	SM, is coupled to relevant SM fermions (denoted as $f$) as follows:
	\begin{equation}
		{\cal L}\supset\sum_{f}g_{f}Z'_{\mu}\overline{f}\gamma^{\mu}f\thinspace.\label{eq:x}
	\end{equation}
	Here the $g_{f}$ should be viewed as effective couplings. 
	They may be fundamental, induced by kinetic or mass mixing (see the discussion in the next section), or generated by loop-level processes (see for instance~\cite{Chauhan:2020mgv}). 
	The SM fermion $f$ can be either chiral (e.g. $f=e_{R}$, $\nu_{L}$,
	$u_{L}$) or non-chiral (e.g. $f=e$, $\mu$, $d$). 
	In principle, one could also introduce an axial coupling when considering non-chiral fermions, however in our phenomenological analysis we will assume that the charged leptons couple purely vectorially to the $Z'$. This provides the most economical solution to the $(g-2)_\mu$ anomaly and is realised in various popular scenarios such as $L_{\mu }-L_{\tau}$. 
	For neutrinos, since light right-handed neutrinos ($\nu_{R}$) with sizeable couplings  would be in conflict with cosmological observations\footnote{More specifically, for a sub-GeV $Z'$ coupled to a light $\nu_{R}$,
		the effective coupling needs to be smaller than $\sim10^{-8}$\textemdash see e.g. Fig.~6
		in Ref.~\cite{Luo:2020fdt}.}, we assume that they are either decoupled from the $Z'$ (in which case they can be light), or sufficiently heavy that they do not contribute
	to the invisible decay width of the $Z'$. 
	Thus, we take $g_{\nu_\alpha} \equiv g_{\nu_{\alpha L}}$ for neutrino flavour $\alpha$.

	We note here that while in this work we treat the $g_{f}$ as independent parameters for different fermions, they may potentially be correlated in specific models that address gauge invariance, electroweak symmetry breaking, and kinetic and mass mixing in detail. To remain maximally
	model-independent, we concentrate on the generic framework proposed
	in Eq.~\eqref{eq:x}. 
	For kinetic and mass 
	mixing, our analysis and results can be readily applied according
	to the discussion in the next subsection.

	\subsection{Kinetic and mass mixing\label{subsec:kinetic}}
	
	In general,  when the SM is extended by an extra $U(1)'$, there is kinetic mixing of the form,
	\begin{equation}
		{\cal L}\supset-\frac{\varepsilon}{2}F^{\mu\nu}F'_{\mu\nu}\thinspace,\label{eq:m-3}
	\end{equation}
	where $F^{\mu\nu}$ and $F'_{\mu\nu}$ are the field strength tensors of the SM $U(1)_{Y}$ and $U(1)'$, with $\varepsilon$ the kinetic mixing parameter. 
	After electroweak symmetry breaking, there may be mass mixing between the gauge boson of the $U(1)'$ and the $Z$-boson of the SM, which can be written as\footnote{Strictly speaking, since the $Z'$ and $Z$ in Eq.~\eqref{eq:m-3a} are
		not physical mass eigenstates, one should differentiate the notation
		in the original basis from that in the mass basis. Here, for
		simplicity, we neglect the difference. For a rigorous treatment, see Refs.~\cite{Lindner:2018kjo,Bauer:2018onh}.
	}
	\begin{equation}
		{\cal L}\supset \sin \theta \, m_{Z}^{2} Z^{\mu}Z'_{\mu}\thinspace,\label{eq:m-3a} 
	\end{equation}
	where $\theta$ is the mass mixing parameter. 
	After necessary transformations to redefine the physical mass eigenstates
	as $Z$ and $Z'$, Eqs.~\eqref{eq:m-3} and \eqref{eq:m-3a} lead to  mixing-induced effective
	couplings of the $Z'$ to all SM fermions, see e.g.~Ref.~\cite{Lindner:2018kjo,Bauer:2018onh}. 
	Here we list the  leading-order results for $f=\ell_{L,R}$
	($\ell=e$, $\mu$, $\tau$) and $\nu_{L}$:
	\begin{align}
		g_{\ell_{L}} & \approx\epsilon e\frac{1-r_{m}/(2c_{W}^{2})}{1-r_{m}}+\frac{1}{2}g_{Z}\left(2s_{W}^{2}-1\right)\sin\theta,\label{eq:x-8}\\
		g_{\ell_{R}} & \approx\epsilon e\frac{1-r_{m}/c_{W}^{2}}{1-r_{m}}+g_{Z}s_{W}^{2}\sin\theta,\label{eq:x-9}\\
		g_{\nu_{L}} & \approx\epsilon e\frac{-r_{m}/(2c_{W}^{2})}{1-r_{m}}+\frac{1}{2}g_{Z}\sin\theta,\label{eq:x-10}
	\end{align}
	where 
	$(s_{W},\ c_{W})\equiv(\sin\theta_{W},\ \cos\theta_{W})$
	with $\theta_{W}$ the Weinberg angle; $g_{Z}\equiv g/c_{W}$ with
	$g$ the $SU(2)_{L}$ gauge coupling of the SM, and $\epsilon$ and $r_{m}$ are defined as 
	\begin{equation}
		\epsilon\equiv\varepsilon c_{W},\ 
		r_{m}\equiv\frac{m_{Z'}^{2}}{m_{Z}^{2}}\,.
		\label{eq:x-15}
	\end{equation}
	In Eqs.~\eqref{eq:x-8}-\eqref{eq:x-10}, the first term comes from kinetic mixing and the second from mass mixing. 
	Note that there are different conventions for the definitions of $\varepsilon$ 
	and $\epsilon$ 
	in the literature. To avoid potential confusion, we have defined both.  
	Our conventions are consistent with e.g. Refs.~\cite{Bauer:2018onh, SHiP:2020vbd}. When fermion $f$ has a nonzero $U(1)'$ charge, $Q_f'$, then one can 
	add $g'Q_f'$ to these effective couplings, where $g'$ is the fundamental $U(1)'$ gauge coupling, as long as both couplings are perturbatively small.\\

	There are some noteworthy limits often considered in the literature\footnote{Note that in these limits, all SM fermions are neutral under the $U(1)'$.}:
	\begin{itemize}
		\item The dark photon limit ($r_{m}\rightarrow0$): When the $Z'$ is very light,
		the kinetic mixing leads to photon-like couplings which are parity-conserving
		for charged fermions and proportional to their electric charges:
		\begin{equation}
			\lim_{r_{m}\rightarrow0,\ \theta\rightarrow0}(g_{e_{L}},\ g_{e_{R}},\ g_{\nu_{L}})=\left(\epsilon e,\ \epsilon e,\ 0\right).\label{eq:x-11}
		\end{equation}
		
		\item Hypercharge limit ($r_{m}\gg1$): When the $Z'$ is very heavy, one can
		see that the $\epsilon$ terms in Eqs.~\eqref{eq:x-8}-\eqref{eq:x-10}
		are proportional to their hypercharges while the $\sin \theta$ term vanishes, thus:
		\begin{equation}
			\lim_{r_{m}\rightarrow\infty,\ \theta\rightarrow0}(g_{e_{L}},\ g_{e_{R}},\ g_{\nu_{L}})\propto\left(1/2,\ 1,\ 1/2\right).
		\end{equation}
		\item $Z$-like  limit ($\epsilon\rightarrow0$): The $\sin\theta$ terms
		in Eqs.~\eqref{eq:x-8}-\eqref{eq:x-10} are proportional to the SM
		$Z$ couplings to the respective fermions. Hence, when $Z'$ is coupled to fermions only via $Z'$-$Z$ mass mixing, the couplings are
		$Z$-like:
		\begin{equation}
			\lim_{\epsilon \rightarrow\infty}(g_{e_{L}},\ g_{e_{R}},\ g_{\nu_{L}})\propto\left( s_{W}^{2}-1/2,\ s_{W}^{2},\ 1/2 \right).
		\end{equation}
	\end{itemize}
	
	\begin{figure}
		\centering
		
		\usetikzlibrary{decorations.pathmorphing,decorations.markings}

		\begin{tikzpicture}

			\node [text=black,rotate=0] at (-3.2,2) {$\mu$};
			\node [text=black,rotate=0] at (-3.2,-1) {$\mu$};

			\begin{scope}[thick,decoration={
					markings,
					mark=at position 0.5 with {\arrow{stealth}}}
				]    
				
				\node [text=black,rotate=0] at (-1,0.8) {$Z'$};
				\draw[postaction={decorate}] (-2,0.5)--(-3,-1);    
				\draw[postaction={decorate}] (-3,2)--(-2,0.5);    
				\draw [decorate,decoration=snake](-2,0.5)--(-0.5,0.5);
			\end{scope}

			\begin{scope}[thick,decoration={
					markings,
					mark=at position 0.5 with {\arrow{stealth}}}
				]  
				
				\draw[postaction={decorate}] (2,0.5)--(1,-1);
				\draw[postaction={decorate}] (1,2)--(2,0.5);
				\draw [postaction={decorate}] (3,0.5) to[out=90,in=90] (4.0,0.5) ;
				\draw [postaction={decorate}] (4.0,0.5) to[in=-90,out=-90]  (3,0.5);	
				\draw [decorate,decoration=snake](4,0.5)--(5,0.5);
				\draw [decorate,decoration=snake](2,0.5)--(3,0.5);
			\end{scope}

			\node [text=black,rotate=0] at (4.8,0.8) {$Z'$};
			\node [text=black,rotate=0] at (2.4,0.8) {$\gamma, Z$};
			\node [text=black,rotate=0] at (1.6,2.2) {$f=e,\nu,q$};
			\node [text=black,rotate=0] at (0.8,-0.8) {$\overline{f}$};
			\node [text=black,rotate=0] at (3.5,1) {$\mu$};
			\node  [text=black,rotate=0,scale=2] at (0.5,0.5) {$\Rightarrow$ };
			
		\end{tikzpicture}
		
		\caption{The loop diagram from which the presence of the $Z'$-$\mu$-$\mu$
			coupling generally implies the existence of $Z'$ couplings to
			other fermions. \label{fig:loop}}
	\end{figure}
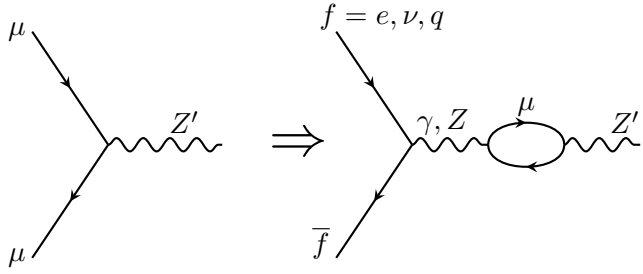
	
	It is important to note that while
	tree-level kinetic and mass mixing could be suppressed by tuning $\epsilon$
	and $\theta$ to sufficiently small values, loop-induced mixing generally
	exists. 
	Assuming that the $Z'$ interacts with muons (in order to explain the $(g-2)_\mu$ anomaly), this generates couplings of the $Z'$ to electrons and
	quarks, as illustrated in Fig.~\ref{fig:loop}. Unless the tree-level mixing is fine-tuned to cancel the loop-induced mixing\footnote{This would violate 't Hooft's technical naturalness~\cite{tHooft:1979rat}.}, generally we expect an approximate lower bound on $g_{f}$ by adapting results in Ref.~\cite{Chauhan:2020mgv} to the loop diagram in Fig.~\ref{fig:loop},
	\begin{equation}
		|g_{f}|\gtrsim\frac{\alpha}{3\pi}|g_{\mu}Q_f|\log\left(\frac{\Lambda^2}{m_{\mu}^2}\right) ,
		\label{eq:m-7}
	\end{equation}
	where $Q_f$ is the electric charge of $f$, $\alpha\equiv e^{2}/(4\pi)$, and $\Lambda$ denotes the new physics scale at which UV divergences are cancelled. 
	In the case of a gauged $L_{\mu}-L_{\tau}$, for instance, which is a
	popular model to address the muon $g-2$ anomaly, the loop diagram of Fig.~\ref{fig:loop} combined with a similar $\tau$ loop is free from UV divergences and leads to the loop-induced couplings~\cite{Araki:2017wyg},
	\begin{equation}
		g_{f}=-\frac{\alpha}{3\pi}g_{\mu}Q_f\log\left(\frac{m_{\mu}^2}{m_{\tau}^2}\right).\label{eq:x-12}
	\end{equation}
	As it has been shown
	in previous studies (see e.g. \cite{Abdullah:2018ykz,Bauer:2018onh,Dev:2020drf}),
	the loop-induced couplings play a crucial role in the $L_{\mu}-L_{\tau}$
	model when confronting it with existing experimental constraints.

	\subsection{Muon $g-2$  and the viable parameter space of muonic $Z'$ models \label{subsec:paraspace}}
	
\begin{figure}
		\centering
		
		\includegraphics[width=0.99\textwidth]{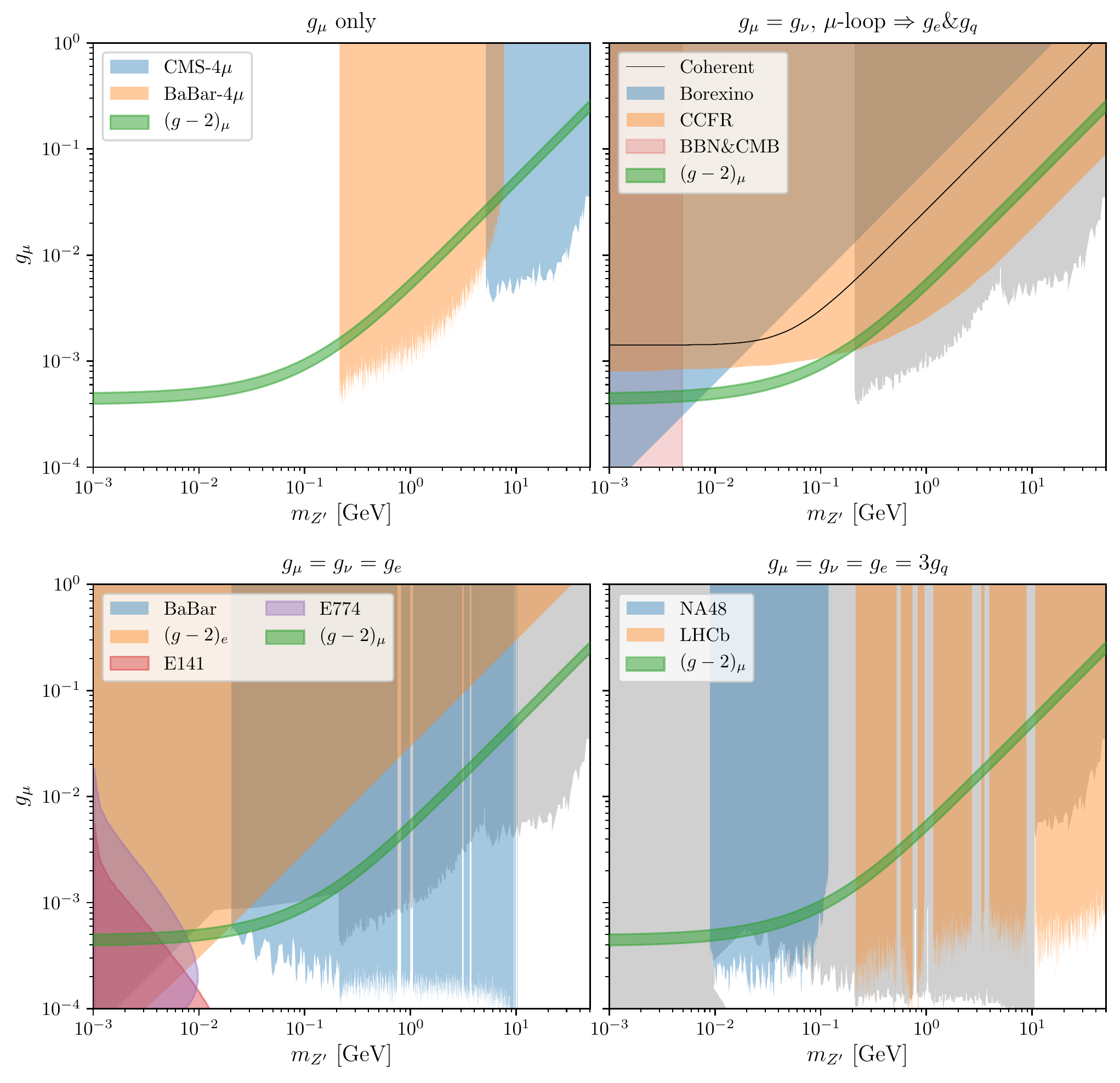}
		
		\caption{\label{fig:current}Current viable parameter spaces for the $Z'$ when
			it couples exclusively to $\mu$ (upper left), couples equally to
			$\mu$ and $\nu_{\mu,L}$ (upper right), couples universally to leptons
			(lower left), or couples to all SM fermions (lower right). The green band reduces the muon $g-2$ anomaly to within 1$\sigma$. Grey bounds correspond to ones already introduced in previous panels.
		}
	\end{figure}

	The muonic coupling, $g_{\mu}$, can be responsible for the discrepancy
	between the SM prediction and experimental measurements of the muon's anomalous magnetic moment.
	The contribution of a $Z'$ can be evaluated using the
	general formula~\cite{Leveille:1977rc}, 
	\begin{equation}
		\Delta a_{\mu}=\frac{g_{\mu}^{2}}{4\pi^{2}}\left(\frac{m_{\mu}}{m_{Z'}}\right)^{2}\int_{0}^{1}\frac{(1-x)x^{2}}{1-x+(m_{\mu}/m_{Z'})^{2}x^{2}}dx\thinspace.\label{eq:m-4}
	\end{equation}
	Here we have assumed that the $Z'$ couples purely vectorially to muons, as mentioned in Sec.~\ref{subsec:generic}. 
	The recent Fermilab measurement of $a_{\mu}$ combined with the previous
	BNL E821 result gives \cite{Muong-2:2006rrc,Muong-2:2021ojo,Aoyama:2020ynm}
	\begin{equation}
		a_{\mu}({\rm Exp})-a_{\mu}({\rm SM})=(25.1\pm5.9)\times10^{-10}\thinspace,\label{eq:m-5}
	\end{equation}
	which indicates a 4.2$\sigma$ deviation. Using the value in Eq.~\eqref{eq:m-5},
	we plot green bands in Fig.~\ref{fig:current} which gives the region where the $Z'$ reduces the tension to within 1$\sigma$. As can be seen, for $m_{Z'}\ll m_{\mu}$, $g_{\mu}$ needs to be around $4.5\times10^{-4}$
	to account for the discrepancy. For $m_{Z'}\gg m_{\mu}$, however, the
	required magnitude of $g_{\mu}$ increases approximately
	linearly with $m_{Z'}$, $g_{\mu}\approx0.05\times(m_{Z'}/10\ {\rm GeV})$. 
	
	Depending on $Z'$ interactions with other SM fermions, there are
	various experimental constraints. If the $Z'$ couples exclusively to
	the muon (though this is theoretically unlikely, especially when taking  gauge invariance and loop corrections into consideration), only collider searches
	for 4$\mu$ final states (e.g. $e^{+}e^{-}\rightarrow\mu^{+}\mu^{-}Z'$
	with $Z'\rightarrow\mu^{+}\mu^{-}$) can constrain it. 
	BaBar \cite{BaBar:2016sci} and CMS \cite{CMS:2018yxg} 4$\mu$ searches have
	excluded the $Z'$ explanation of the anomaly for $m_{Z'}>2m_{\mu}$, as shown in the upper
	left panel of Fig.~\ref{fig:current}. 
	
	From the point of view of gauge invariance, given a coupling
	to $\mu$, the $Z'$ is likely also to have couplings to $\nu_{\mu L}$ with
	$g_{\nu_{\mu}}$ comparable to $g_{\mu_{L}}$.\footnote{It might be possible that $Z'$ only couples to right-handed leptons,
		so that this argument does not apply.} When the $Z'$ has a neutrino coupling, neutrino scattering data can
	be used to constrain it. There are various important bounds, including from $\nu_{\mu}+e^{-}$
	(Borexino \cite{Bellini:2011rx,Bauer:2018onh}) and $\nu_{\mu}+N$ (COHERENT \cite{Abdullah:2018ykz})
	elastic scattering\footnote{Another $\nu_{\mu}$ elastic scattering experiment, CHARM II,
		provides the most restrict bound for $m_{Z'}\gtrsim100$ MeV among
		$\nu_{\mu}$ elastic scattering experiments, but as we have checked,
		in this regime it is weaker than the CCFR bound\textemdash see e.g.~\cite{Bilmis:2015lja,Lindner:2018kjo}.}, and $\nu_{\mu}+N\rightarrow\nu_{\mu}+N+\mu^{+}+\mu^{-}$ trident
	scattering (CCFR \cite{Altmannshofer:2014pba}), shown in the upper right panel under the assumption $g_{\nu_\mu} = g_\mu$.
	The neutrino elastic scattering bounds are derived from one-loop processes that involve a muon loop which connects $Z'$ to $\gamma$, and hence to electrons and nucleons.
	The trident scattering corresponds to opening the loop,
	which leads to two muons in the final state.

	When $m_{Z'}$ is $\mathcal{O}(\text{MeV})$, cosmological constraints from Big Bang Nucleosynthesis (BBN) and Cosmological Microwave Background (CMB) are also relevant.
	The bounds depend on the $Z'$ coupling to neutrinos and to electrons. Ref.~\cite{Sabti:2019mhn} studied the bound on $m_{Z'}$ as a function of the electron to neutrino annihilation ratio. 
	For our purposes, this refers to the ratio of the decay rates into electrons and neutrinos:
	\begin{eqnarray}
		\frac{\Gamma(Z' \to e^+ e^-)}{\Gamma(Z' \to \nu^+ \nu^-)} \approx \frac{2g_e^2}{g_\nu^2}\, ,
	\end{eqnarray}
	with the factor of 2 since we sum over final state spins, where we define
	\begin{equation}
		g_\nu^2 \equiv g_{\nu_e}^2 + g_{\nu_\mu}^2 + g_{\nu_\tau}^2 \, .
	\end{equation}
	We take the BBN+Planck row of table VI in Ref.~\cite{Sabti:2019mhn} and interpolate between the data points to obtain the bounds. Depending on this ratio, it varies from 1.3 to 9.4 MeV. For the upper right panel of Fig.~\ref{fig:current}, using the loop-induced coupling given by Eq.~\eqref{eq:x-12} for $g_e$, we find that the bound is $m_{Z'} \gtrsim 4.9$ MeV, although it should be kept in mind that this number has an $\mathcal{O}(1)$ uncertainty. 
	There is also a bound on $Z'$ couplings to neutrinos due to white dwarf cooling \cite{Dreiner:2013tja}. 
	However, since the constraint is very similar to the one obtained from Borexino but relatively more uncertain, it is not included in the plots.

	In the lower panels of Fig.~\ref{fig:current}, we impose further constraints assuming the presence of tree-level electron and quark couplings. In this case, collider experiments (e.g. BaBar, NA48, LHCb) are able to
	search directly for a $Z'$ resonance. Moreover, electron $g-2$ and
	beam dump (e.g. E774, E141)\footnote{
		Other beam dump experiments such as E137 and Orsay are less restrictive in the region of parameter space we consider.} 
	experiments provide complementary constraints. These constraints
	were obtained by employing the {\tt DARKCAST} package \cite{Ilten:2018crw}.

	Fig.~\ref{fig:current} demonstrates that in order to accommodate the
	muon $g-2$ anomaly, the $Z'$ boson needs to have suppressed
	couplings to electrons and light quarks in comparison to $g_{\mu}$, i.e.
	\begin{equation}
		|g_{\mu}|\gtrsim\{|g_{e}|,\ |g_{\nu}|,\ |g_{u}|,\ |g_{d}|\}\thinspace.\label{eq:m-6}
	\end{equation}
	From now on, Eq.~\eqref{eq:m-6} will be considered as a guiding principle when we are concerned with experimental constraints.

	\section{Sensitivity of future BD experiments \label{sec:BD}}
	
	Future beam dump (BD) experiments such as SHiP and SeaQuest have great potential to probe a light, weakly-interacting $Z'$, provided that
	its invisible decay width is not too large. In the present framework, where we do not introduce any new fermions, only neutrinos contribute to the invisible decay width. Therefore, the
	BD sensitivity depends largely on the strength of neutrino couplings, $g_{\nu}$. 
	
	To illustrate the influence of $g_{\nu}$, we start in Sec.~\ref{subsec:simplified} with a simplified
	scenario where only $g_{e}$ and $g_{\nu}$ are present,
	and then generalise in Sec.~\ref{subsec:hadro} to more realistic situations where heavy leptons
	($\mu$, $\tau$) and hadronic states are also taken into account. Then in Sec.~\ref{subsec:casestudy} we perform case studies for SHiP, FASER, and SeaQuest.

	\subsection{Influence of neutrino couplings: a simplified scenario \label{subsec:simplified}}
	
	In the presence of only $g_{e}$ and $g_{\nu}$ couplings, the $Z'$ can be produced at the target in BD experiments via the bremsstrahlung
	process, $e^{-}+N\rightarrow e^{-}+N+Z'$, with the production cross
	section proportional to $g_{e}^{2}$,
	\begin{equation}
		\sigma_{{\rm prod}}\propto g_{e}^{2}\thinspace.\label{eq:x-1}
	\end{equation}
	Once produced, the $Z'$ particle decays during its flight and causes
	an observable signal only if it penetrates the shielding material
	(of length $L_{{\rm sh}}$) and decays to visible states within the fiducial
	decay region (of length $L_{{\rm dec}}$).  The probability
	of the $Z'$ flying through the shielding and decaying in the fiducial
	decay region is
	\begin{equation}
		P=e^{-L_{{\rm sh}}/L_{{\rm flight}}}\left(1-e^{-L_{{\rm dec}}/L_{{\rm flight}}}\right),\label{eq:x-3}
	\end{equation}
	where the mean flight distance, $L_{{\rm flight}}$, is given by
	\begin{equation}
		L_{{\rm flight}}=\frac{\tau_{0}v}{\sqrt{1-v^{2}}},\ v=\frac{p_{Z'}}{\sqrt{m_{Z'}^{2}+p_{Z'}^{2}}}\thinspace.\label{eq:x-2}
	\end{equation}
	Here $v$, $p_{Z'}$, and $\tau_{0}$  denote the velocity, momentum, and lifetime (at rest) of the $Z'$ particle, respectively.
	
	Decaying in the fiducial decay region is not enough. To be detectable,
	the $Z'$ must decay to visible final states ($e^{+}e^{-}$, $\mu^{+}\mu^{-}$,
	or hadronic states). This is taken into account by including
	the branching ratio of visible decays,  $\text{BR}_{{\rm vis}}$. Thus,
	the event rate at the detector is given by
	\begin{equation}
		R\propto\sigma_{{\rm prod}}\cdot P\cdot\text{BR}_{{\rm vis}}\thinspace.\label{eq:x-4}
	\end{equation}
	
	Since in this simplified scenario the $Z'$ only decays to electrons 
	and neutrinos, 
	$\text{BR}_{{\rm vis}}$ and $\tau_{0}$ are given by
	\begin{align}
		\text{BR}_{{\rm vis}} & =\frac{2g_{e}^{2}}{g_{\nu}^{2}+2g_{e}^{2}}\thinspace,\label{eq:x-5}\\
		\tau_{0}^{-1} & =\Gamma_{Z'\rightarrow\nu\overline{\nu}}+\Gamma_{Z'\rightarrow e\overline{e}}\approx\frac{m_{Z'}}{12\pi}\left(\frac{1}{2}g_{\nu}^{2}+g_{e}^{2}\right),\label{eq:x-6}
	\end{align}
	where we have neglected the electron mass. 
	
	Substituting Eqs.~\eqref{eq:x-5} and \eqref{eq:x-6} into Eq.~\eqref{eq:x-4},
	we obtain $R$ as a function of $g_{e}$, $g_{\nu}$ and $m_{Z'}$.
	The absolute magnitude of $R$  depends on the exposure time, beam
	energy and luminosity, target material, etc. 
	To avoid these experiment-dependent details, we make the replacement $\sigma_{{\rm prod}}\rightarrow g_{e}^{2}$
	in Eq.~\eqref{eq:x-4} and define a dimensionless event rate,
	\begin{equation}
		\overline{R}\equiv g_{e}^{2}\cdot P\cdot\text{BR}_{{\rm vis}}\thinspace.\label{eq:m}
	\end{equation}
	
	Assembling the above pieces, we obtain
	\begin{equation}
		\overline{R}=\frac{g_{e}^{4}}{g_{e}^{2}+g_{\nu}^{2}/2}e^{-\lambda_{{\rm sh}}(g_{e}^{2}+g_{\nu}^{2}/2)}\left[1-e^{-\lambda_{{\rm dec}}(g_{e}^{2}+g_{\nu}^{2}/2)}\right],\label{eq:x-13}
	\end{equation}
	where
	\begin{equation}
		\lambda_{{\rm sh}/{\rm dec}}\equiv L_{{\rm sh}/{\rm dec}}\frac{m_{Z'}^{2}}{12\pi p}\thinspace.\label{eq:x-14-1}
	\end{equation}
	\begin{figure}
		\centering
		
		\includegraphics[width=0.46\textwidth]{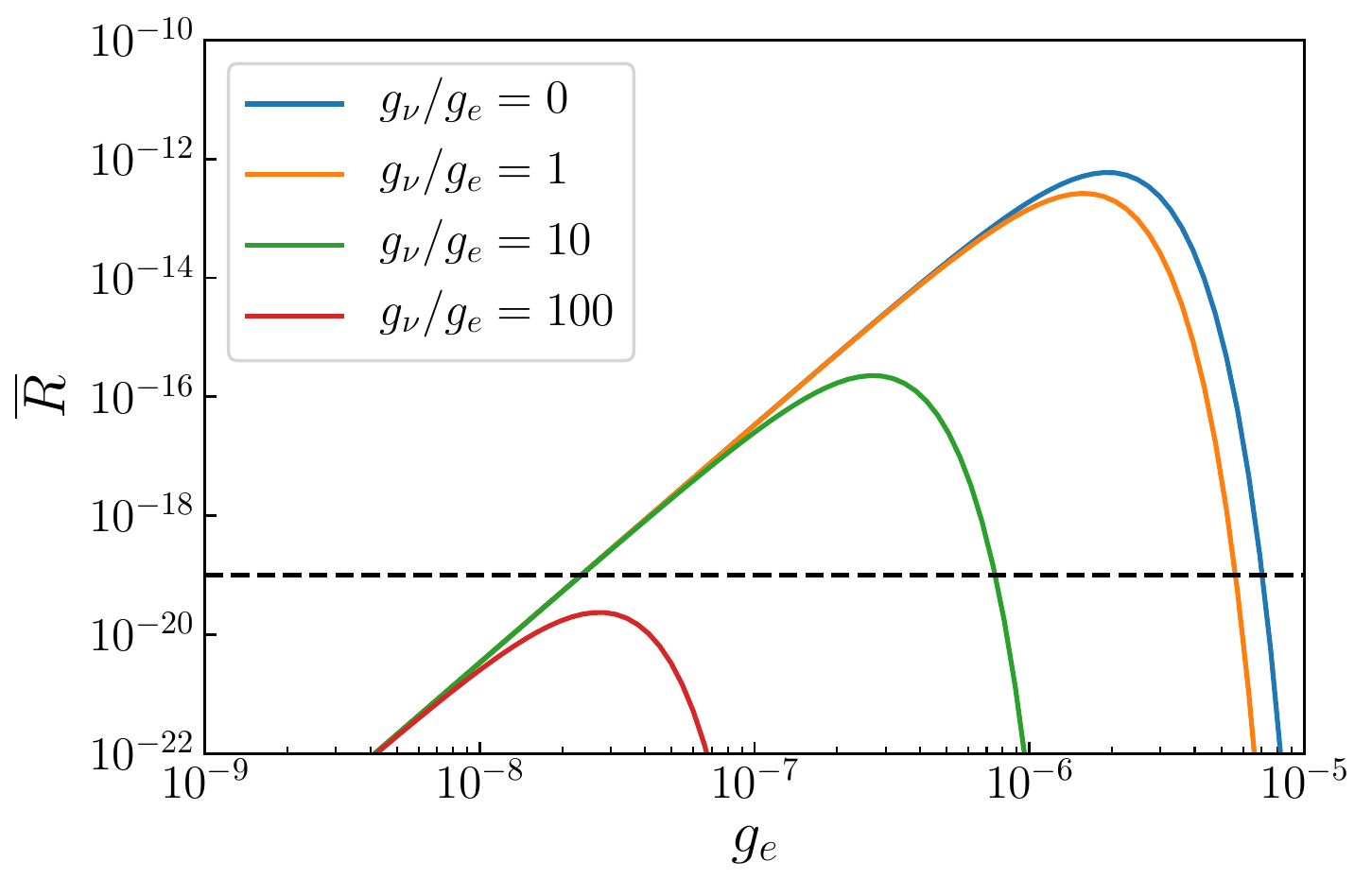}\includegraphics[width=0.45\textwidth]{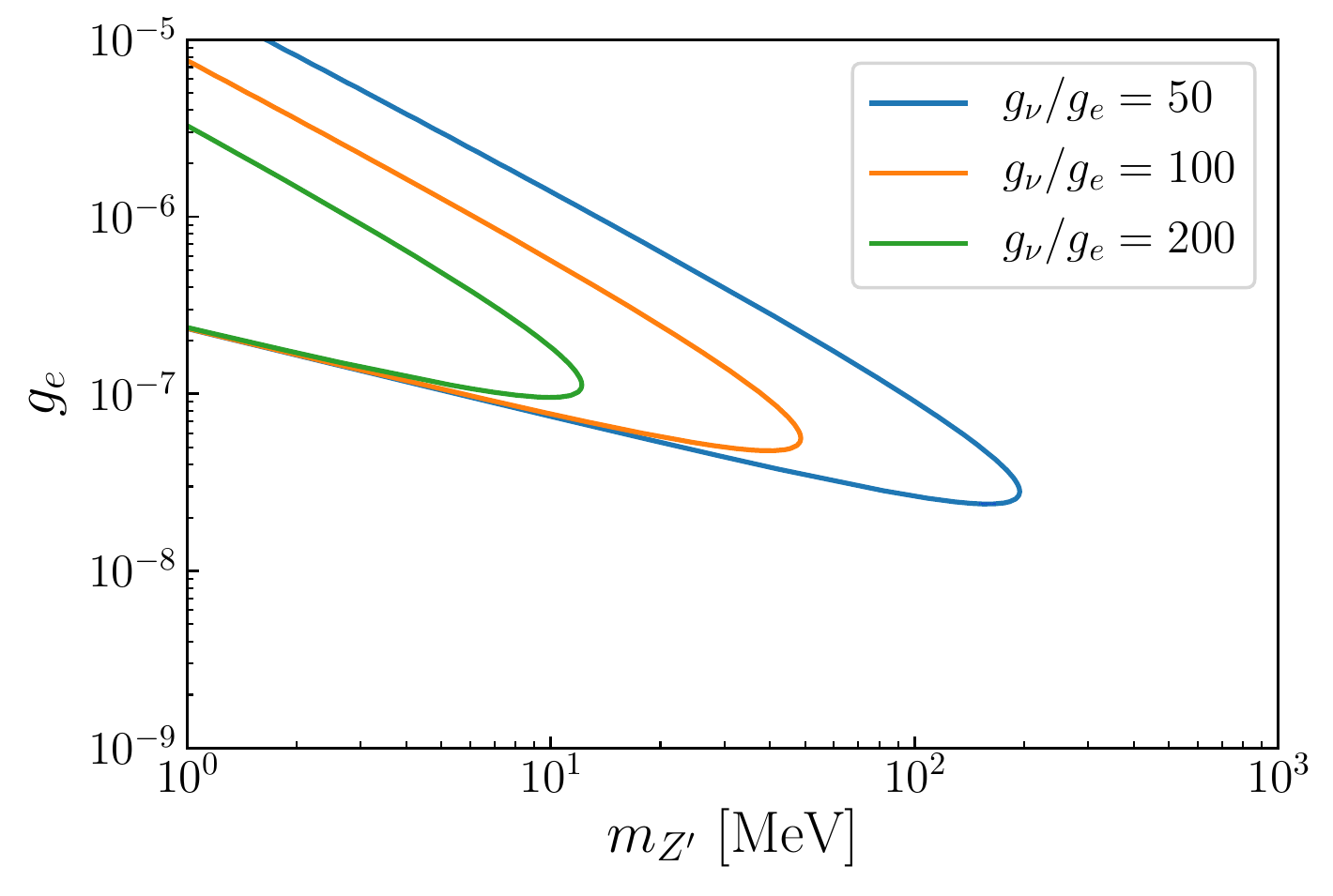}
		
		\caption{Left: the dimensionless event rate $\overline{R}$ (defined in Eq.~\eqref{eq:x-13})
			as a function of $g_{e}$ and $g_{\nu}$, fixing $m_{Z'}=100$ MeV.
			The black horizontal line represents a hypothetical experimental
			sensitivity at $\overline{R}=10^{-19}$ for the purpose of illustration.
			Right: contours of $\overline{R}(m_{Z'},g_{e})=10^{-19}$, demonstrating the different hypothetical experimental sensitivities to $m_{Z'}$ and $g_{e}$.
			\label{fig:R-ge}}
	\end{figure}
	
	In the left panel of Fig.~\ref{fig:R-ge}, we plot $\overline{R}$-$g_{e}$
	curves with $g_{\nu}/g_{e}$ fixed at a few given values. We took $p_{Z'}=200$ GeV, $L_{{\rm sh}}=60\ {\rm m}$, $L_{{\rm dec}}=50\ {\rm m}$, similar to the specifications for SHiP \cite{Alekhin:2015byh}. For illustration, we plot a black dashed
	line at $\overline{R}=10^{-19}$ as a hypothetical experimental sensitivity.
	Typically, a BD experiment is sensitive to $g_{e}$ in a certain
	interval $g_{e}\in[g_{{\rm lower}},\ g_{{\rm upper}}]$ where $g_{{\rm lower}}$
	and $g_{{\rm upper}}$ are determined by the intersection of
	the $\overline{R}$-$g_{e}$ curve with the black dashed line. 
	
	The shape of these curves can be understood as follows. When $g_{e}$
	is sufficiently small, BD experiments lose sensitivity because (i)
	the $Z'$ production rate is suppressed, and (ii) the probability of the $Z'$
	decaying in the fiducial region is suppressed ($\tau_{0}$ and $L_{{\rm flight}}$
	are too large, so most decays happen after the $Z'$ has flown further than $L_{{\rm sh}}+L_{{\rm dec}}$). 
	On the other hand, when $g_{e}$ is too large the BD experiments also lose sensitivity due to small $L_{{\rm flight}}$, in which case most of the particles decay in the shielding. 
	As can be seen from the left panel of Fig.~\ref{fig:R-ge}, the size of $g_{\nu}/g_{e}$ can significantly affect $g_{{\rm upper}}$
	but has less impact on $g_{{\rm lower}}$. As $g_{\nu}/g_{e}$ increases, $g_{{\rm upper}}$ decreases because $\text{BR}_{{\rm vis}}$ is reduced. When $g_{{\rm upper}}$ approaches $g_{{\rm lower}}$, the experiment quickly loses sensitivity.

	In the right panel of Fig.~\ref{fig:R-ge}, we plot contours of $\overline{R}=10^{-19}$
	in the $m_{Z'}$-$g_{e}$ plane. If $\overline{R}=10^{-19}$ is interpreted
	as the experimental sensitivity, then the experiment will be able
	to probe the regions enclosed by these contours. As is shown, when
	$g_{\nu}/g_{e}$ increases, these contours shrink not only vertically
	but also horizontally. This implies that larger invisible decay widths
	of the $Z'$ will not only lead to smaller intervals of $[g_{{\rm lower}},\ g_{{\rm upper}}]$, but will also reduce the experimental sensitivity to a heavy $Z'$.

	\subsection{Including hadronic and heavy leptonic states\label{subsec:hadro}}
	
	The simplified analysis above grasps the main features of BD constraints, such as the existence of $g_{{\rm lower}}$ and $g_{{\rm upper}}$, and the effect
	of $g_{\nu}$. In more realistic situations, additional
	decay channels (in particular hadronic states) and the dependence
	of $\sigma_{{\rm prod}}$ on $m_{Z'}$ need to be taken into account. 
	We turn to this now.
	
	For $m_{Z'}>2m_{\mu}\approx212$ MeV, the decay mode $Z'\rightarrow\mu\overline{\mu}$ opens up. 
	For even higher masses, the $Z'$ can also decay to hadronic states.
	Including heavy leptonic final states is straightforward. In the case of $Z'\rightarrow\mu\overline{\mu}$,
	the partial decay width is given by (see e.g. \cite{SHiP:2020vbd})
	\begin{equation}
		\Gamma_{Z'\rightarrow\mu\overline{\mu}}=\frac{g_{\mu}^{2}}{12\pi}m_{Z'}\sqrt{1-\frac{4m_{\mu}^{2}}{m_{Z'}^{2}}}\left(1+\frac{2m_{\mu}^{2}}{m_{Z'}^{2}}\right).\label{eq:m-2}
	\end{equation}
	For $Z'$ masses above about $3.5$ GeV, the $Z'\rightarrow\tau\overline{\tau}$ decay
	mode could also be relevant, and can be included using Eq.~\eqref{eq:m-2}
	with $m_{\mu}\rightarrow m_{\tau}$ and $g_{\mu}\rightarrow g_{\tau}$.

	The widths of possible hadronic decay modes ($Z'\rightarrow{\rm hadrons}$)
	can be computed using the hadron-to-muon cross section ratio of $e^{+}e^{-}$
	collisions, 
	\begin{equation}
		{\cal R}(\sqrt{s})=\frac{\sigma(e^{+}e^{-}\rightarrow\gamma^{*}\rightarrow{\rm hadrons})}{\sigma(e^{+}e^{-}\rightarrow\gamma^{*}\rightarrow\mu\overline{\mu})}\thinspace,\label{eq:m-8}
	\end{equation}
	where $\sqrt{s}$ is the the center-of-mass energy. The ${\cal R}$
	ratio has been well determined by $e^{+}e^{-}$ collision data for
	$0.3\ {\rm GeV}\lesssim\sqrt{s}\lesssim200\ {\rm GeV}$.\footnote{See Fig.~52.2 of \cite{Zyla:2020zbs} or Fig.~2 of \cite{Ilten:2018crw}. 
		The data is available from \url{https://pdg.lbl.gov/2021/hadronic-xsections/}}  Given the experimentally measured ${\cal R}$ values, the total
	hadronic decay width of $Z'$  is given by 
	\begin{equation}
		\Gamma_{Z'\rightarrow{\rm had.}}=\left(\frac{g_{q}}{eQ_{q}}\right)^{2}\Gamma_{\gamma^{*}\rightarrow\mu\overline{\mu}}{\cal R}(m_{Z'})\thinspace,\label{eq:m-9}
	\end{equation}
	where $Q_{q}$ is the electric charge of quark $q$ and we have assumed
	that $Z'$-quark couplings are photon-like so that $g_{q}/Q_{q}$
	are independent of the type of quarks. This is the case for the well-studied
	dark photon scenario, see the dark photon limit in Sec.~\ref{subsec:kinetic}.
	For more general couplings, if hadronic decays are subdominant compared
	to $Z'\rightarrow\mu\overline{\mu}$,  one can still neglect the
	difference between $g_{u}/Q_{u}$ and $g_{d}/Q_{d}$.  The virtual
	photon decay width $\Gamma_{\gamma^{*}\rightarrow\mu\overline{\mu}}$
	can be computed using Eq.~\eqref{eq:m-2} with $g_{\mu}^{2}\rightarrow e^{2}$.
	
	In the presence of decays to hadronic ($m_{Z'}>2m_{\pi}$) and heavy
	leptonic ($m_{Z'}>2m_{\mu}$) states, the $\text{BR}_{{\rm vis}}$
	of Eq.~\eqref{eq:x-5} is modified to
	\begin{equation}
		\text{BR}_{{\rm vis}}=1-\frac{\Gamma_{\nu}}{\Gamma_{\nu}+\Gamma_{\ell}+\Gamma_{{\rm had.}}}\thinspace,\label{eq:m-10}
	\end{equation}
	where $\Gamma_{\nu}$, $\Gamma_{\ell}$, and $\Gamma_{{\rm had}.}$
	are the decay widths of $Z'\rightarrow\nu\overline{\nu}$, $Z'\rightarrow\ell\overline{\ell}$
	($\ell=e$, $\mu$, $\tau$), and $Z'\rightarrow{\rm hadrons}$, respectively.
	The extra decay modes also modifies the lifetime $\tau_{0}$ in Eq.~\eqref{eq:x-6} to 
	\begin{equation}
		\tau_{0}=(\Gamma_{\nu}+\Gamma_{\ell}+\Gamma_{{\rm hadrons}})^{-1}.\label{eq:m-11}
	\end{equation}
	It is also important to note that when $Z'$ is heavy, the production
	cross section becomes dependent on $m_{Z'}$.   Therefore, for SHiP-like
	experiments, we modify Eq.~\eqref{eq:m} as follows,
	\begin{equation}
		\overline{R}\equiv g_{p}^{2}\cdot\frac{\sigma_{{\rm prod}}}{\sigma_{{\rm prod}}^{\star}}\cdot P\cdot\text{BR}_{{\rm vis}}\thinspace,\label{eq:m-12}
	\end{equation}
	where $g_{p}$ denote the effective coupling of $Z'$ to the proton,
	$\sigma_{{\rm prod}}$ can be obtained from e.g.~Fig.~4 in Ref.~\cite{SHiP:2020vbd}\footnote{Ref.~\cite{SHiP:2020vbd} adopted two approaches to evaluate the
		production cross section. One included the standard dipole form factor, while the other took into account the possibility of nuclear resonance
		enhancement, referred to as the vector meson dominance (VMD) model.
		The former leads to a more conservative result than the later, although
		the difference becomes significant only for $m_{Z'}\gtrsim500$ MeV.
		In this work, we adopt the cross section obtained from the former
		approach. } and $\sigma_{{\rm prod}}^{\star}\equiv\sigma_{{\rm prod}}(m_{Z'}=100\ {\rm MeV})$. 
	
	\begin{figure}
		\centering
		
		\includegraphics[width=0.49\textwidth]{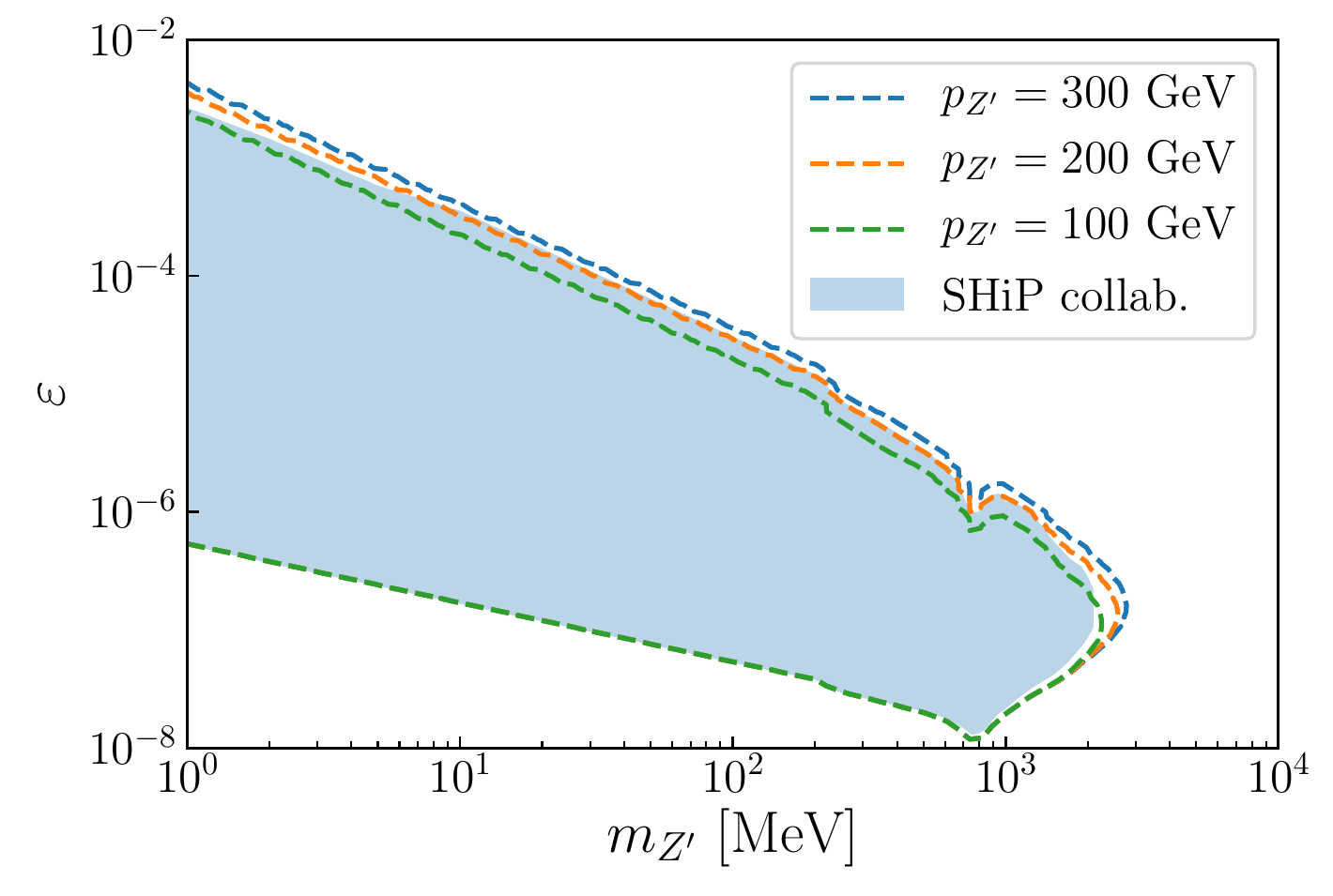}\includegraphics[width=0.49\textwidth]{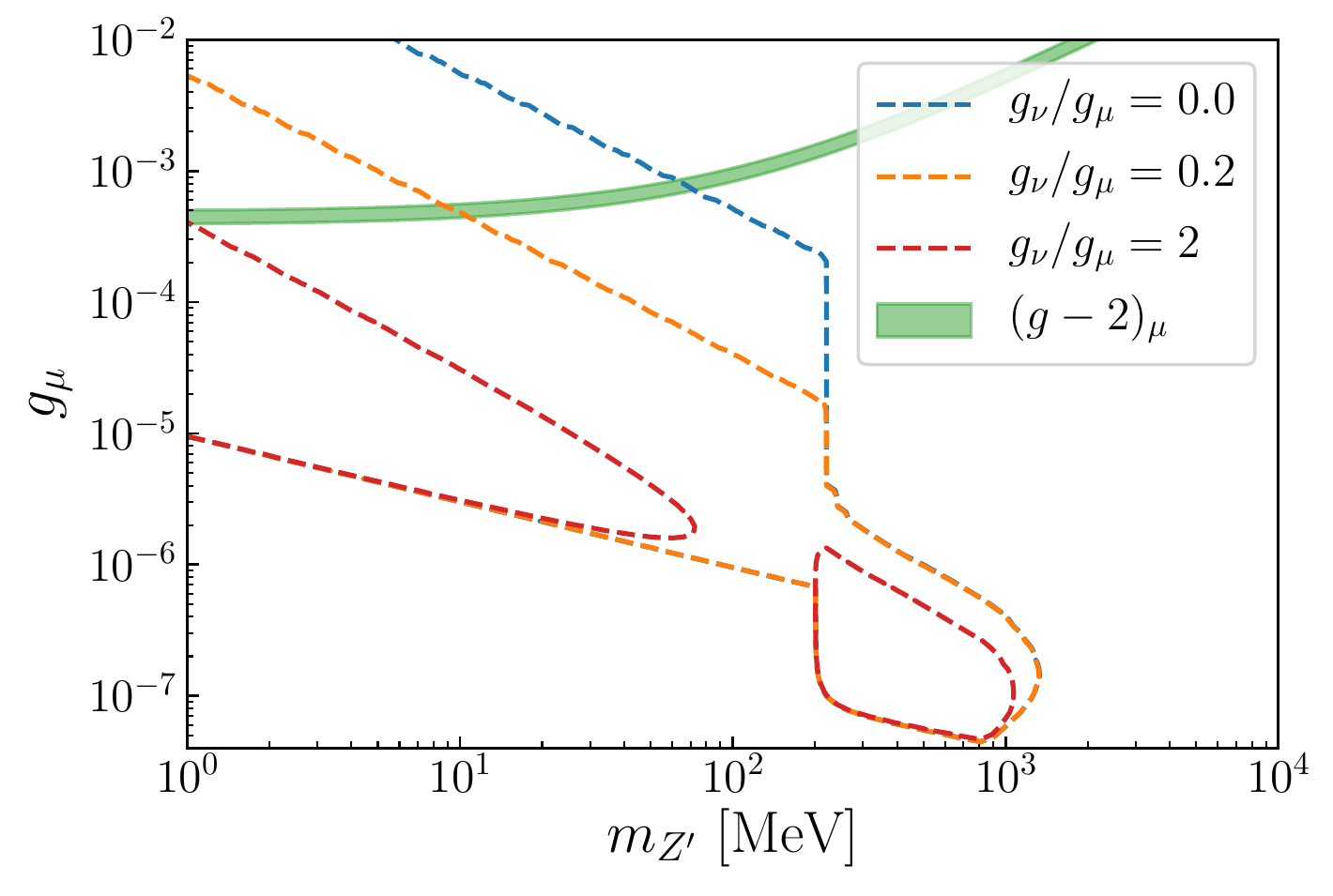}
		
		\caption{Left: Reproduced bounds (dashed) on $\varepsilon$ 
			compared to the bound published by the SHiP collaboration \cite{SHiP:2020vbd} (blue shaded). Right: Bounds on the muonic coupling, $g_{\mu}$, due to the invisible decay of $Z'$.
			The green band in the right panel is the same as in Fig.~\ref{fig:current}, and couplings of $Z'$ to the electron and quarks are assumed at the loop-induced level given by Eq.~\eqref{eq:x-12}.
			\label{fig:Recast}}
	\end{figure}
	
	\subsection{Case studies for SHiP, SeaQuest, and FASER \label{subsec:casestudy}}
	In the left panel of Fig.~\ref{fig:Recast}, we apply Eq.~\eqref{eq:m-12}
	to reproduce the SHiP sensitivity curves for the dark photon scenario
	with $g_{f}=\varepsilon c_{W}eQ_{f}=\epsilon eQ_{f}$. The blue shaded
	region represents the result published by the SHiP collaboration (taken
	from the upper panel in Fig.~13 of Ref.~\cite{SHiP:2020vbd}) and
	the dashed curves are produced by extracting contours of $\overline{R}(m_{Z'},\ \varepsilon)$,
	which are fixed at certain values so that the resulting lower bounds
	at $m_{Z'}=100$ MeV match the SHiP result ($\varepsilon=5.28\times10^{-8}$).
	As indicated in the plot, the momentum $p$ is fixed at several values
	below the proton beam energy ($400$ GeV). 
	We take $L_{{\rm sh}}=60\,{\rm m}$ and $L_{{\rm dec}}=50\,{\rm m}$ for the shielding and decay lengths, following Refs.~\cite{Alekhin:2015byh,SHiP:2020vbd}. 
	The specific values of $\overline{R}$ for the three contours (blue, orange, green) to match $\varepsilon=5.28\times10^{-8}$ at $m_{Z'}=100$ MeV are
	$(0.5,\ 0.75,\ 1.5)\times10^{-20}$, respectively. 
	
	The plot shows that Eq.~\eqref{eq:m-12} can rather accurately describe
	the experimental sensitivity in the dark photon scenario. 
	The choice of the specific value of $p_{Z'}$ does not affect the result significantly, as long as it is a significant fraction of the incoming proton energy. 
	This allows us to recast the SHiP bound on the dark photon to give bounds on a $Z'$ with invisible decays. 
	
	In the right panel of Fig.~\ref{fig:Recast}, we set $p_{Z'}=200$ GeV
	and $\overline{R}=0.75\times10^{-20}$ (corresponding to the orange
	curve in the left panel) 
	to generate the SHiP bound on a $Z'$ with $g_{\nu}=(0,\ 0.2,\ 2)g_{\mu}$. In addition, we assume $g_{p}$ and $g_{e}$ are loop-induced (see Fig.~\ref{fig:loop}) using  Eq.~\eqref{eq:x-12}, which gives $g_p=-g_e$. 
	Modifying this by an $\mathcal{O}(1)$ amount will not qualitatively change our conclusions.

	As is shown in Fig.~\ref{fig:Recast}, the value of $g_{\nu}$
	significantly affects the SHiP sensitivity to $Z'$. 
	For $g_{\nu}=0$, since there is no invisible decay, the $Z'$ can only
	decay to electrons when $m_{Z'}$ is below $212$ MeV. 
	When it is above $212$ MeV, the decay mode $Z'\rightarrow\mu\overline{\mu}$
	opens and, due to the large coupling, the $Z'$ lifetime drastically
	decreases, leading to a very substantial drop in the upper and lower
	bounds on $g_\mu$. For $g_{\nu}=0.2g_{\mu}$, the high-mass regime ($>212$ MeV) is not significantly affected because $Z'\rightarrow\mu\overline{\mu}$
	is still the dominant decay mode. In the low-mass regime, however, the
	branching ratio of invisible decay is enhanced, leading to a reduction
	of the upper bound by an order of magnitude. When $g_{\nu}$ further increases, both high-
	and low-mass regimes are affected. In particular, when $g_\nu/g_\mu = 2$, the sensitivity region is divided into two separate regions: $m_{Z'}\lesssim 80$ MeV and $m_{Z'}\gtrsim 212$ MeV.
	
	\begin{figure}[t]
		\centering    
		\includegraphics[width=0.49\textwidth]{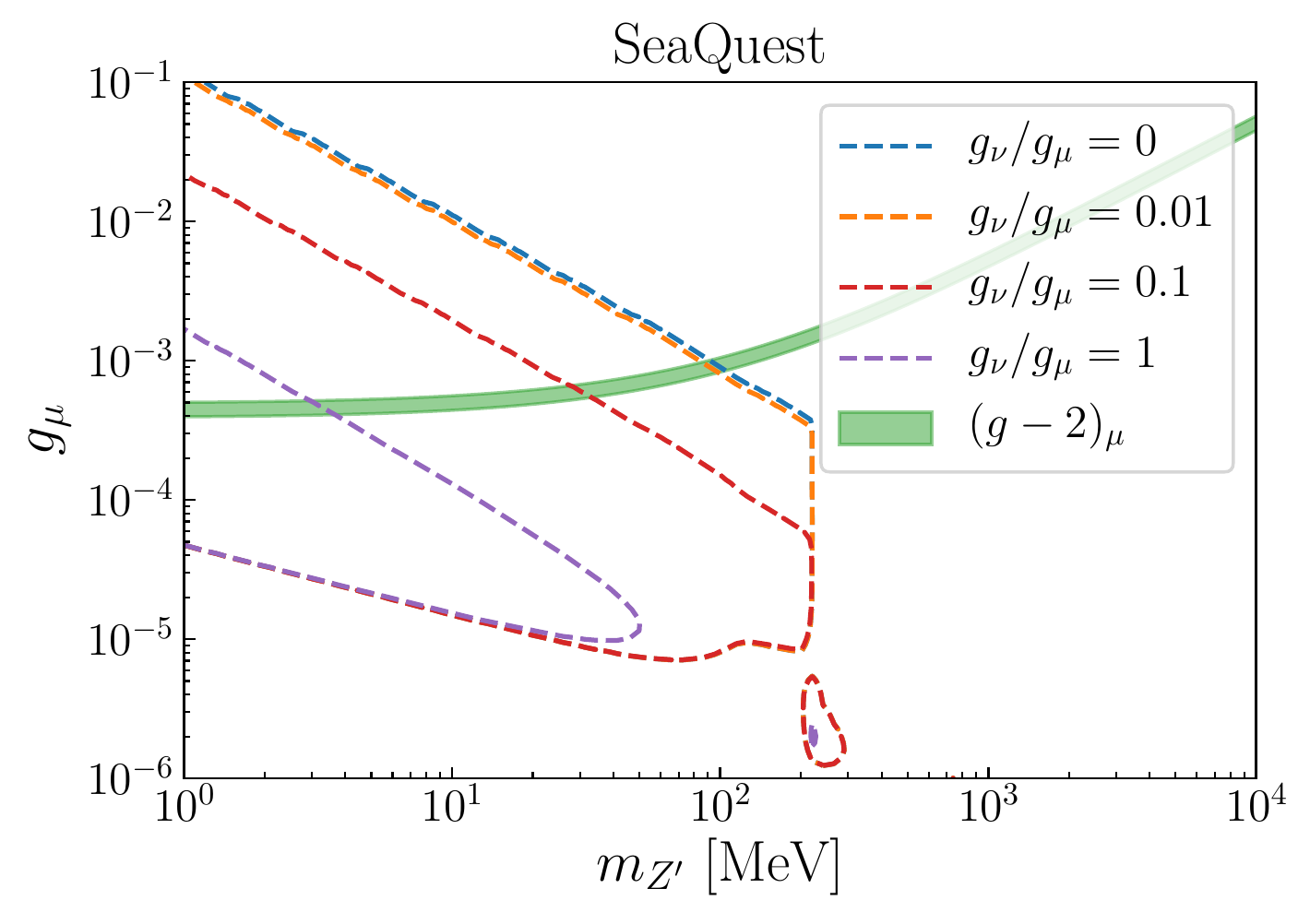}\includegraphics[width=0.49\textwidth]{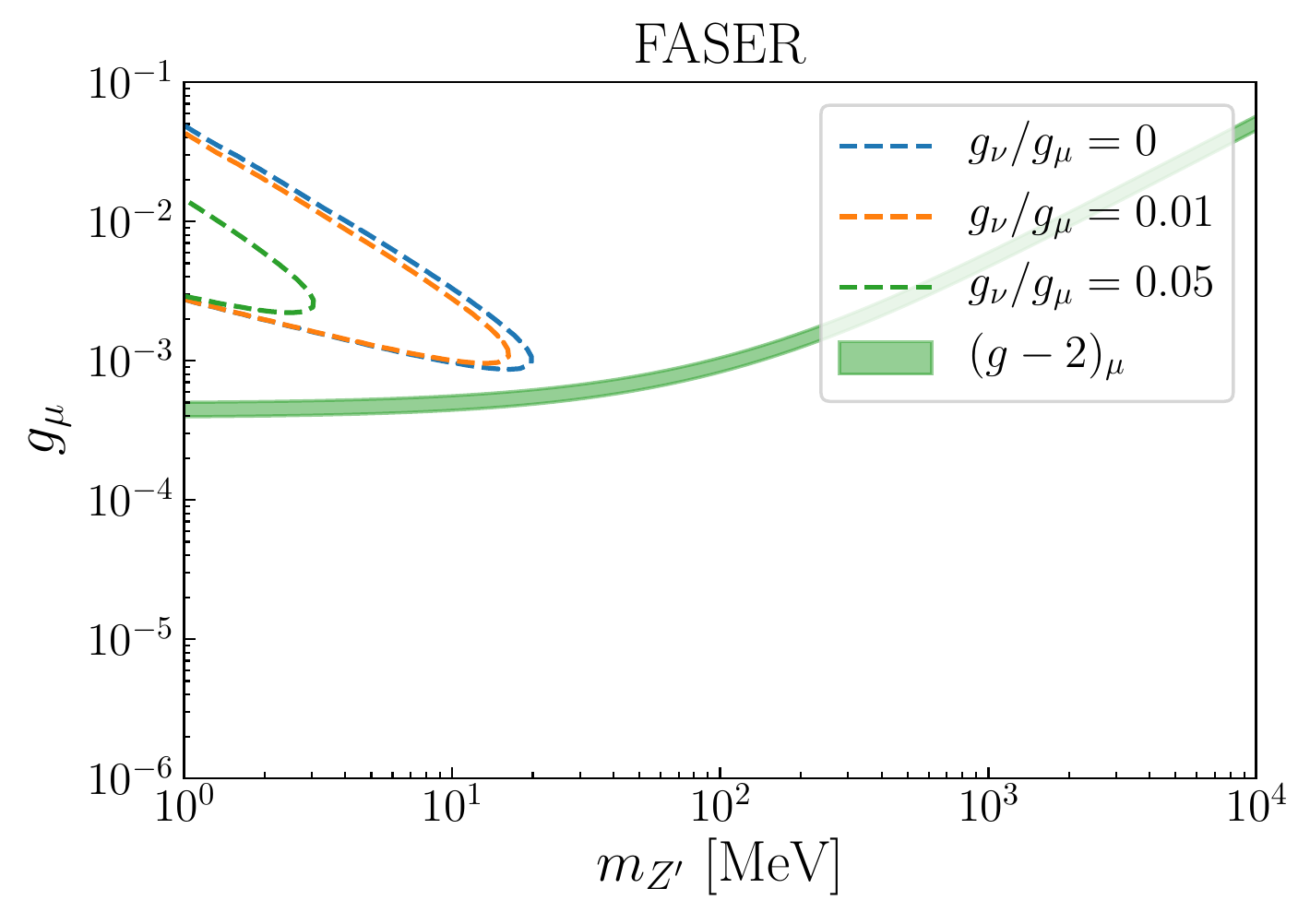}    
		\caption{Similar to the right panel of Fig.~\ref{fig:Recast}, this time for SeaQuest (left) and FASER (right).
			\label{fig:Recast2}}
	\end{figure}
	
	\begin{table*}
		\caption{\label{tab:exp} Configurations of future BD experiments}
		\centering %
		\begin{tabular}{ccccc}
			\hline \hline 
			Experiments & $L_{{\rm sh}}$ {[}meter{]} & $L_{{\rm dec}}$ {[}meter{]} & proton beam energy & Ref.\tabularnewline
			\hline 
			SHiP & 60 & 50 & 400 GeV & \cite{SHiP:2020vbd}\tabularnewline
			SeaQuest & 5 & 1 & 120 GeV & \cite{Berlin:2018pwi}\tabularnewline
			FASER & 480 & 5$\times3$  & 7 TeV & \cite{FASER:2018eoc}\tabularnewline
			\hline \hline 
		\end{tabular}
		
	\end{table*}

	The above analysis can be straightforwardly adapted to other similar BD experiments such as FASER and SeaQuest. We summarize their configurations in Tab.~\ref{tab:exp} and perform similar analyses. For SeaQuest, we adopt the result in Fig.~2 of Ref.~\cite{Berlin:2018pwi} to determine the mass dependence of the production rate. The FASER experiment is based on 7+7 TeV proton collision at the LHC, and thus technically not a BD experiment. Nevertheless, we can treat it as a BD experiment because it is sensitive to $Z'$ masses below 3 GeV, where proton bremsstrahlung and meson decays are the dominant processes for $Z'$ production. Using Eq.~\eqref{eq:m-12}, we successfully reproduce the anticipated dark photon results in both Ref.~\cite{Berlin:2018pwi} (SeaQuest) and Ref.~\cite{FASER:2018eoc} (FASER). 
	Like the above analysis for SHiP, we also find that the dependence of the results on $p_{Z'}$ is weak and the best fit is obtained when $p_{Z'}$ is set at half the proton beam energy. 
	
	In Fig.~\ref{fig:Recast2}, we recast the published dark photon bounds of SeaQuest and FASER to the bounds on $Z'$. For SeaQuest, we present the sensitivity reach of its Phase-I run with $1.44\times 10^{18}$ protons on target (POT). The result could be further improved by its Phase-II run with $10^{20}$ POT. Due to the uncertainty of the experimental configuration for its Phase II, we only include Phase I of SeaQuest in our analysis. For FASER, there are also two proposed configurations, with the integrated luminosity $\int dL$ and the decay volume $L_{\rm dec}$ given by $(\int dL,\ L_{\rm dec})=(150\ {\rm fb}^{-1},\ 1.5 \times 3\ {\rm m})$  and $(3\ {\rm ab}^{-1},\ 5 \times 3\ {\rm m})$---see Sec.~II~E in \cite{FASER:2018eoc}. We adopt the latter and find that even with the enhanced integrated luminosity, the FASER sensitivity still cannot reach the green $(g-2)_\mu$ band in Fig.~\ref{fig:Recast2}. This is mainly due to the low collision rate of collider-based experiments compared to fixed-target experiments ($3\ {\rm ab}^{-1}$ only corresponds to $2.3 \times 10^{17}$\ POT).  


	\section{Combined results and discussions}
	\label{sec:results}
	
	\begin{figure}
		\centering
		
		\includegraphics[width=0.49\textwidth]{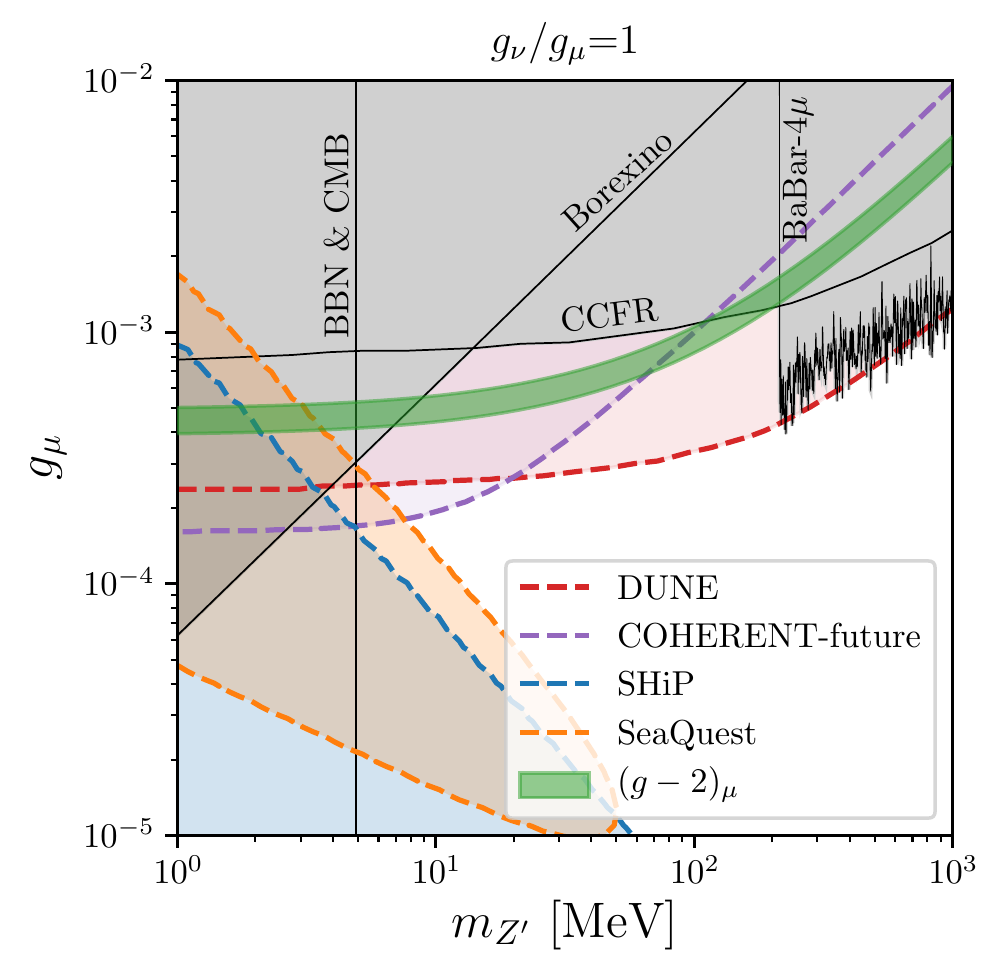}\includegraphics[width=0.49\textwidth]{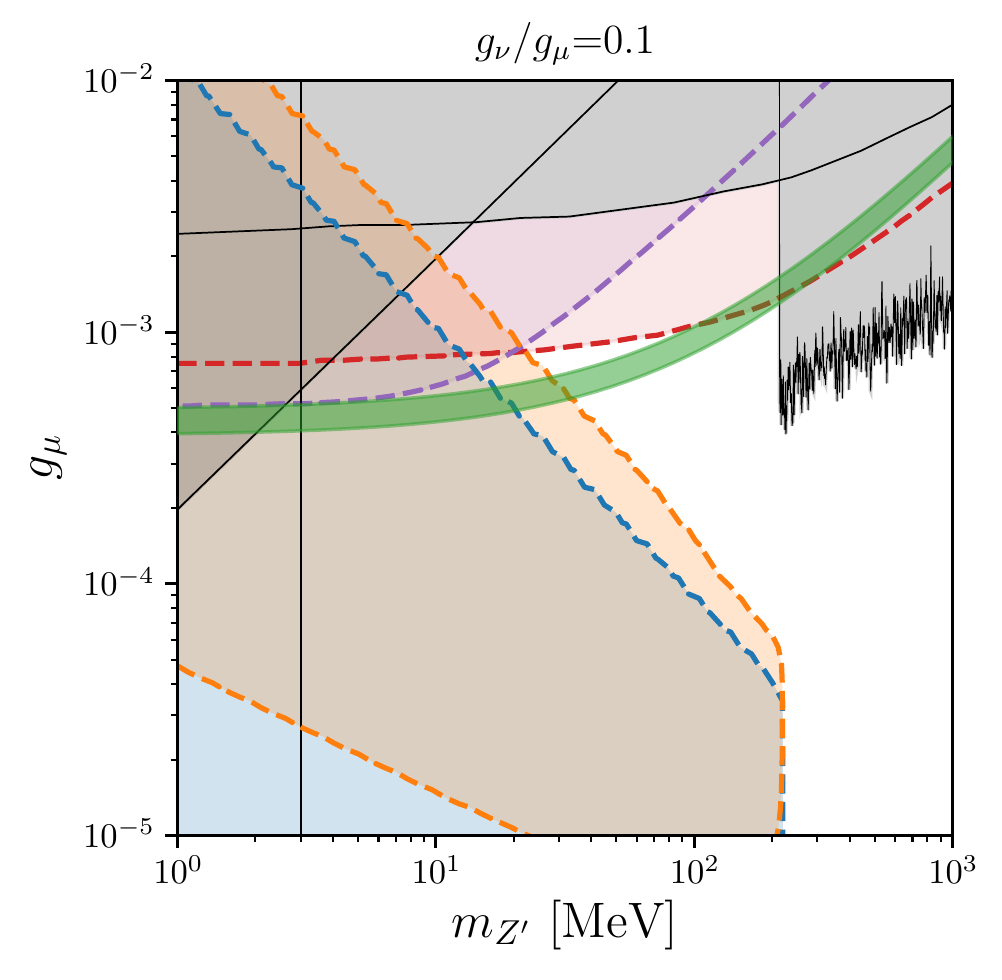}
		
		\includegraphics[width=0.49\textwidth]{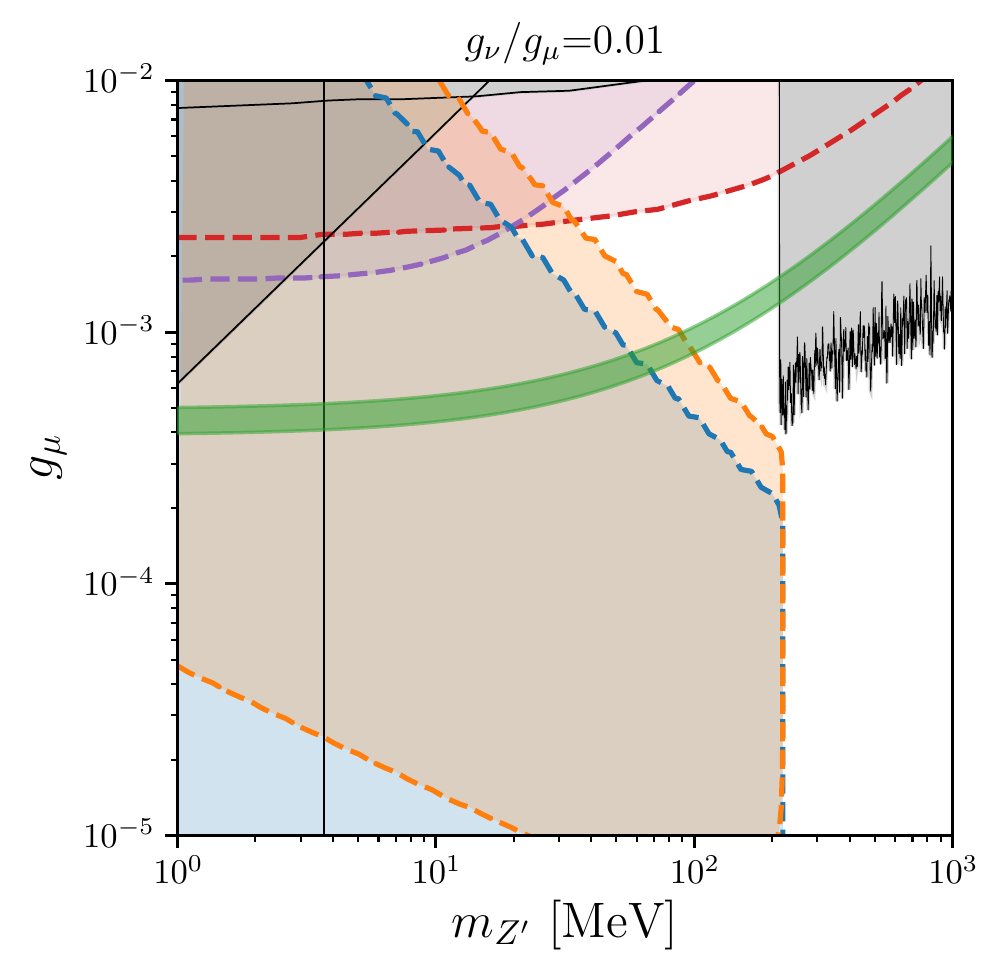}\includegraphics[width=0.49\textwidth]{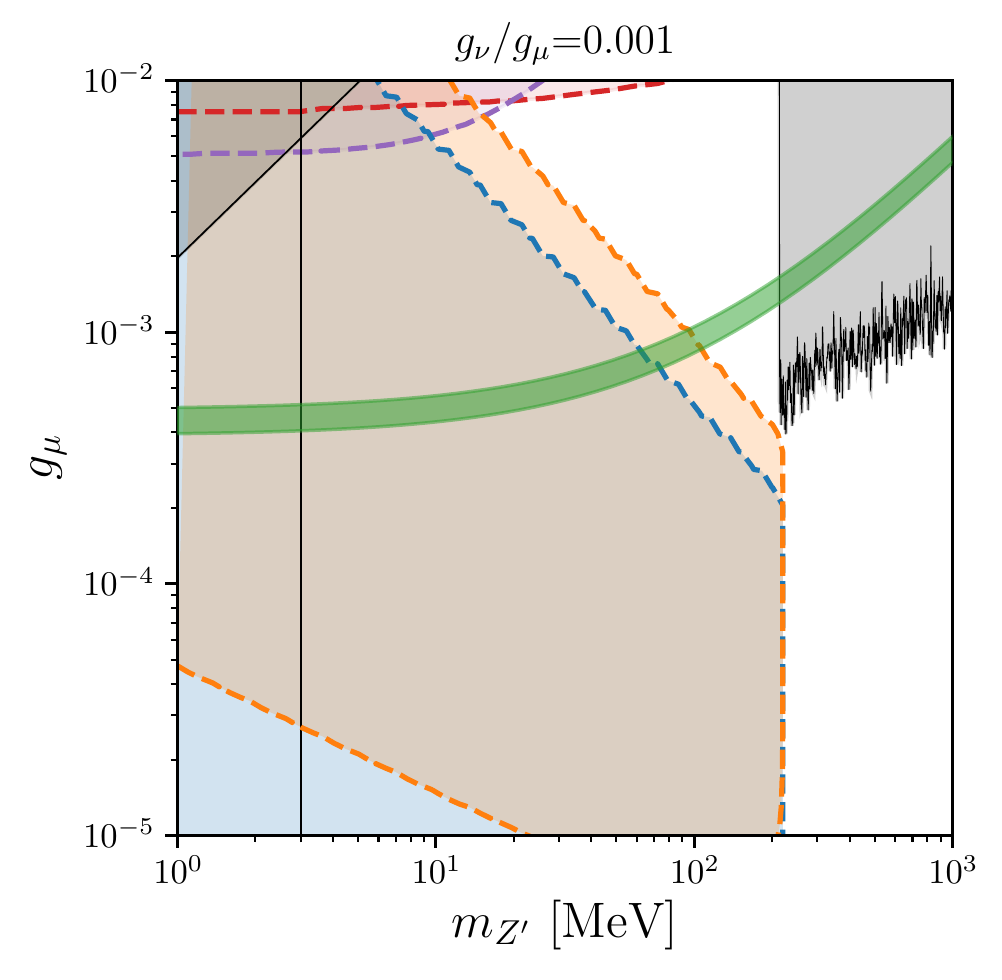}
		
		\caption{\label{fig:four-plot} Prospects of future BD and neutrino experiments probing the $Z'$ solution to the muon $g-2$ anomaly. The upper left panel assumes $g_\nu/g_{\mu}=1$ and states the different bounds in the figure, the other panels with $g_\nu/g_{\mu}=$0.1, 0.01, and 0.001 employ the same conventions. Couplings of $Z'$ to the electron and quarks are assumed to be at the loop-induced level given by Eq.~\eqref{eq:x-12}.
		}
	\end{figure}
	
	Using the results obtained in the previous sections, in Fig.~\ref{fig:four-plot} we present the prospects of probing the $Z'$ as a solution to the muon $g-2$ anomaly in future BD experiments. 
	We show SHiP and SeaQuest sensitivity curves for $g_\nu/g_{\mu}=$1, 0.1, 0.01, and 0.001. FASER results are absent due to their weak sensitivity\textemdash see the discussion in Sec.~\ref{sec:BD}. 
	Existing limits including BaBar-4$\mu$, Borexino, CCFR, and BBN\&CMB  are presented as solid black curves and their details have been explained in Sec.~\ref{subsec:paraspace}.

	As we have discussed, the neutrino coupling $g_{\nu}$ plays a crucial role here because the invisible decay width can substantially weaken the sensitivity of BD experiments. 
	On the other hand, a sizeable $g_{\nu}$ can be constrained by various neutrino scattering experiments. When $g_\nu$ increases, neutrino scattering bounds are stronger. When $g_\nu$ decreases, BD experiments provide more restrictive bounds.
	Therefore, constraints from BD and neutrino experiments are complementary to each other.\footnote{In this paper, we took $g_\nu = g_{\nu_\mu}$, and therefore $g_{\nu_e} = g_{\nu_\tau} = 0$, in order to directly compare bounds from BDs with those from muon neutrino scattering experiments. Introducing non-zero $g_{\nu_e}$ and $g_{\nu_\tau}$ would modify this, however it would introduce a range of additional bounds e.g. the stringnet bound on electron neutrino scattering from TEXONO \cite{Deniz:2009mu}.}

	To show this complementarity, we adapt results from previous studies on neutrino trident scattering\footnote{
		Although $\nu$-$e$ elastic scattering at the DUNE near detector can also be sensitive to the $Z'$, in the low-mass limit it is about a factor of two weaker in terms of $g_{\mu}$ than the trident scattering.}
	(similar to CCFR) at the DUNE near detector~\cite{Ballett:2019xoj} and coherent elastic neutrino-nucleus scattering (CE$\nu$NS) at future COHERENT detectors~\cite{Abdullah:2018ykz}.  
	The DUNE curves presented in Fig.~\ref{fig:four-plot} assume a 75 tonne LAr near detector with 5-year data taking for each of $\nu$ and $\overline{\nu}$ modes, and a 5\% normalization uncertainty. The COHERENT curves assume 10 tonne$\cdot$year exposure of NaI and Ar detectors with the current neutrino flux from the 
	Spallation Neutron Source. 
	When $g_{\nu}/g_{\mu}$ varies, the neutrino scattering bounds are rescaled by a factor of $\sqrt{g_{\nu}/g_{\mu} }$.
	
	As is shown in Fig.~\ref{fig:four-plot}, for $g_\nu/g_\mu=1$, future neutrino scattering experiments such as DUNE or COHERENT will be fully able to probe or exclude the $Z'$ solution to muon $g-2$. Reducing $g_{\nu}$ can significantly alleviate neutrino scattering bounds but in this case SHiP and SeaQuest will provide rather restrictive constraints. For $g_\nu/g_\mu$ varying from 1 to 0.001, the combination of future neutrino and BD experiments can generally probe most of the current viable parameter space of the $Z'$ solution to the muon $g-2$ anomaly.

	\section{Conclusions}
	\label{sec:con}
	A light $Z'$ is a simple and popular explanation for the muon $g-2$ anomaly. 
	Within this class of solutions, in order to evade stringent experimental bounds,  the $Z'$ couplings to electrons and quarks have to be suppressed. 
	In this work, we impose general lower bounds on these suppressed couplings by taking into account possible loop corrections, and study the prospect of probing such a $Z'$ in future BD experiments. 
	After introducing the formalism and current experimental status in Sec.~\ref{sec:Framework}, we investigate in detail the sensitivity of future beam dump experiments in Sec.~\ref{sec:BD}. 
	When the $Z'$ coupling to neutrinos is suppressed with respect to its coupling to muons, these beam dumps will have the capacity to rule out\textemdash or discover\textemdash a large portion of the successful $Z'$ parameter space, despite the smallness of its coupling to electrons and quarks. 
	For $g_\nu/g_\mu \lesssim 0.01$, SHiP and SeaQuest will rule out $Z'$ explanations of the anomaly with $m_{Z'} \lesssim 100$ MeV, leaving only a fairly narrow window given that $m_{Z'} \gtrsim 2m_\mu$ is already excluded by BaBar 4$\mu$. 
	Models with larger $Z'$ couplings to neutrinos somewhat circumvent these bounds, but are constrained by current or future neutrino scattering experiments. 
	There is thus a powerful complementarity between beam dump and neutrino scattering experiments, as outlined in Sec.~\ref{sec:results} and displayed in Fig.~\ref{fig:four-plot}. 
	The $g_\nu = g_\mu$ case, which arises for instance in the $L_{\mu }-L_{\tau}$ model, will be completely covered by the experiments we considered.

	The muon $g-2$ anomaly is particularly interesting not just because it is an indication of new physics, but because the nature of the anomaly suggests that the new physics, if it is exists, may very well be discovered in the near future. 
	This works highlights some promising avenues for such a potential discovery.

	\section*{Acknowledgments}
	This work is supported by IISN convention No. 4.4503.15, by the ``Probing dark matter with neutrinos'' ULB-ARC convention, and by the F.R.S./FNRS under the Excellence of Science (EoS) project No. 30820817 - be.h ``The $H$ boson gateway to physics beyond the Standard Model''. 
	R.C. thanks the UNSW School of Physics, where he is a Visiting Fellow, for their hospitality during this project.

	\bibliographystyle{JHEP}
	\bibliography{ref}
	
\end{document}